# DARK MATTER SEARCH IN THE EDELWEISS EXPERIMENT

M. Chapellier[a*], A. Benoit[b], L. Bergé[c], A. Broniatowski[c], B. Chambon[d],
G. Chardin[e], P. Charvin[c,f], M. De Jésus[d], P. Di Stefano[e], D. Drain[d], L. Dumoulin[c],
J. Gascon[d], G. Gerbier[e], C. Goldbach[g], M. Goyot[d], M. Gros[e], J.P. Hadjout[d],
S. Hervé[e], A. Juillard[e], A. de Lesquen[e], M. Loidl[e], J. Mallet[e], S. Marnieros[c],
O. Martineau[d], N. Mirabolfathi[c], L. Miramonti[e], L. Mosca[c], X.-F. Navick[e],
G. Nollez[g], P. Pari[a], C. Pastor[d], E. Simon[d], M. Stern[d], L. Vagneron[d]

[a]*CEA, Centre d'Etudes Nucléaires de Saclay, DSM/DRECAM, F-91191 Gif-sur-Yvette Cedex, France*

[b]*Centre de Recherche sur les Très Basses Températures, BP 166, 38042 Grenoble, France*

[c]*CSNSM, IN2P3-CNRS, Univ. Paris XI, bat. 108, F-91405 Orsay Cedex, France*

[d]*IPN Lyon and UCBL, IN2P3-CNRS, 43 Bd. du 11 novembre 1918, F-69622 Villeurbanne Cedex, France*

[e]*CEA, Centre d'Etudes Nucléaires de Saclay, DSM/DAPNIA, F-91191 Gif-sur-Yvette Cedex, France*

[f]*Laboratoire Souterrain de Modane, CEA-CNRS, 90 rue Polset, F-73500 Modane, France*

[g]*Institut d'Astrophysique de Paris, INSU-CNRS, 98 bis Bd. Arago, F-75014 Paris, France*

[*]*E-mail chap@spec.saclay.cea.fr*

Preliminary results obtained with 320g bolometers with simultaneous ionization and heat measurements are described. After a few weeks of data taking, data accumulated with one of these detectors are beginning to exclude the upper part of the DAMA region. Prospects for the present run and the second stage of the experiment, EDELWEISS-II, using an innovative reversed cryostat allowing data taking with 100 detectors, are briefly described.

## 1   Introduction

The EDELWEISS experiment is a WIMP direct detection experiment set in the Fréjus Underground Laboratory, adjacent to a highway tunnel connecting France and Italy under the Alps. The rock overburden reduces the muon background flux by a factor $\approx 2 \ 10^6$ while the neutron flux, originating mainly from the surrounding rock, is $\approx 4 \ 10^{-6}$ neutron/cm$^2$/s [1].

EDELWEISS has developed cryogenic germanium detectors with simultaneous measurement of charge and phonon signals [2]. A first motivation for the development of these detectors lies in the possibility to lower the energy threshold compared to classical detectors [3]. A second motivation comes from the fact that present experiments appear to be mostly limited by the radioactive background rate of the detectors and by the systematic uncertainties of the rejection scheme using pulse shape discrimination (PSD) techniques in NaI crystals [4]. The simultaneous measurement of the charge and phonon signal using cryogenic detectors, or light and phonon signals, allows a much more reliable discrimination between the main source of radioactive background, producing electron recoils, and the nuclear recoils expected from WIMP interactions (see e.g. [5]).

In the following, we discuss the effective rejection factor obtained with a new 320g bolometer with guard ring and aluminum electrodes.

## 2 The 320g bolometer.

This 320 g germanium bolometer has been described in Ref. [6]. It is equipped with aluminum electrodes, used as a possible solution to reduce charge collection problems in surface events with respect to implanted electrodes. This detector benefits from a guard ring identifying signals occurring near the edges of the detector. At present, two 320g Ge bolometers have been tested. The complete set of three 320 g bolometers should be installed (fig. 3) at the end of the present run.

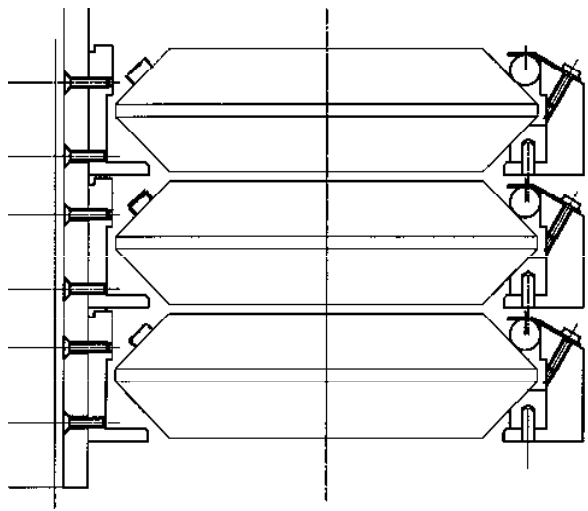

**Figure 1**: Geometrical arrangement of the EDELWEISS "1kg" stage. Each 320g detector (diam.70mm, thickness 20mm) is laying on Teflon pieces. The space between the detectors is approx. 1.5 mm.

On the lower 320 g Ge detector, both central and guard ring electrodes are polarized at 6.3 volts. A linear cross-talk between the two channels (see fig 2) has been observed, but this relatively limited effect can be accounted for by the simultaneous analysis of signals recorded on both electrodes. A fiducial volume, conservatively estimated to 50% of the detector volume from neutron calibrations, is defined by selecting events with notable amplitude on the central electrode only. The temperature increase due to an interaction is measured through a DC-polarised NTD germanium sensor, with a resistance around 4MΩ. The temperature is stabilized within ±10 μK around 27 mK.

The ionization channel resolutions are 2 keV FWHM baseline and 2.8keV @ 122keV for the central zone, with a threshold of ≈5 keV e.e., mostly limited by microphonics, and 1.1 keV baseline and 1.8keV @ 122keV for the guard ring, with a threshold of ≈3 keV e.e. The heat channel exhibits a resolution of 2.2 keV baseline and 3keV @ 122keV.

The raw data trigger rate, prior to software analysis, is of the order of a few events per minute.

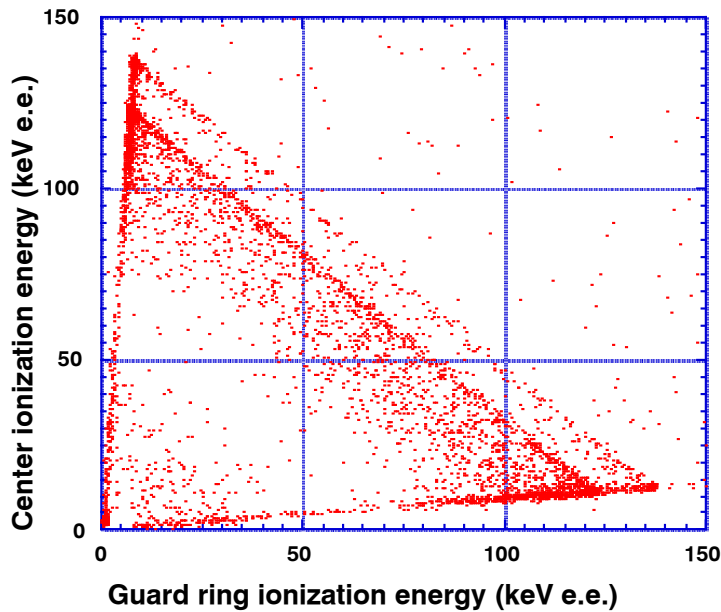

**Figure 2**: Scatter diagram of the ionization amplitudes of the center region versus that of the guard ring. The points on the nearby horizontal line are pure guard events, those on the nearly vertical line are pure center events. Events with signals detected on both channels lie between the two lines. These calibration data have been obtained using a removable [57]Co source.

### 3 Preliminary results

We present here the preliminary results obtained in a physics data taking of 6.3 kg × day, and a fiducial volume exposure of 3.1 kg × day. On Figure 3, the histogram of the ionization amplitude/recoil energy ratio —the quenching factor—, normalized to 1 for electron recoils, has been represented. A paraffin shield of 30cm surrounding the cryostat has been used to reduce the fast neutron flux from the rock by more than two orders of magnitude. In this figure, the histogram of the quenching factor obtained using two previous heat/ionization bolometers is also represented for comparison. It can readily be seen that the performance increase is important. In particular, by restricting the data to the central region only, no events are observed in the [30, 100] keV recoil energy interval over the ±2σ neutron zone, which corresponds to a nuclear recoil acceptance of ≈0.95 for the central part of the detector. A conservative exclusion contour in the WIMP mass vs. cross-section scatter diagram has been derived from these data, and is represented in Fig. 4. It can be seen that the upper part of the region associated with the DAMA region [7] is excluded. From the absence of any recorded event in the [30, 100] keV recoil energy, a significant sensitivity increase over the next few months can reasonably be expected. It is important to note that this limit is obtained without any neutron background subtraction.

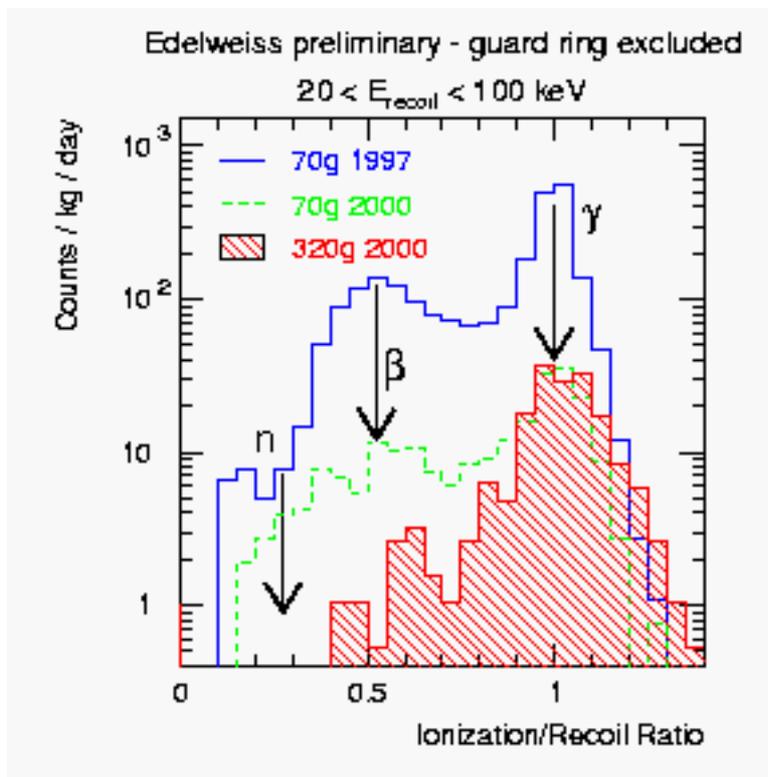

**Figure 3**: Histogram of quenching factor for data recorded with : (upper curve) our first 70g Ge detector, (dashed curve) a second generation 70g Ge detector and (lower curve) using the present 320g detector (fiducial volume). The three curves are normalized to a rate per *kg day*

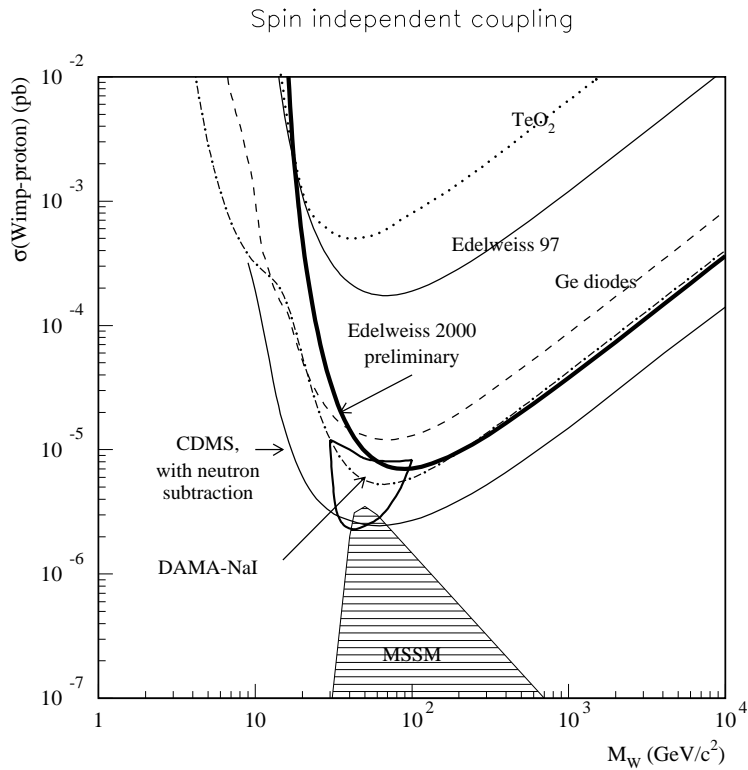

**Figure 4**: Exclusion limit (dark solid curve) obtained using the present data of the fiducial volume of the 320g bolometer. No events are observed in the ±2 σ neutron zone in the [30, 100] keV recoil energy interval using a 3.1 kg × data sample. The quoted exposure takes into account efficiency and fiducial volume corrections. The limits reported in Refs. [7], [8] and [9] have also been represented.

## 4 Perspectives and conclusions.

With an effective mass of 160 gram, ≈600 smaller than the DAMA detector, and with an exposure ≈10 000 times smaller, the present detector already partially excludes the DAMA candidate [7]. In a forthcoming run, we will install three 320 g Ge detectors similar to the present one. Data takings realized during the present "1 kg" stage are expected to last until the end of 2001 and should reach the sensitivity required by the most favorable SUSY models. Improvements on the wiring and the heat channel should also allow to lower the energy threshold by a factor ≈2.

To explore more deeply the parameter space of SUSY models, it is clear, from the absence of any candidate event after ≈6 weeks of data taking, that a larger number of detectors will be required. Therefore, we are actively preparing a second stage of the experiment, EDELWEISS-II, which will use a 100-liter dilution fridge of novel geometry, represented on Fig. 5 and presently built in CRTBT Grenoble. This large cryostat, able to accommodate 100 detectors and their electronics, is expected to be installed in the Fréjus Underground Laboratory in 2002. This second stage is expected to provide an increase in sensitivity by more than two orders of magnitude over the best present performances.

**Acknowledgements**

The help of the technical staff of the Fréjus Underground Laboratory and the participating laboratories is gratefully acknowledged. This work has been partially funded by the EEC Network program under contract ERBFMRXCT980167.

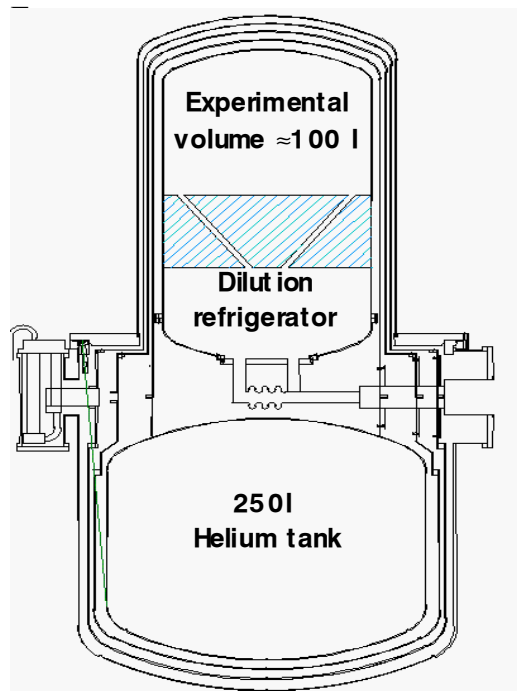

**Figure 5**: Reversed dilution cryostat of EDELWEISS-II. The hatched part represents a lead or copper shield used to protect the experimental volume (upper part) from the dilution refrigerator. The available experimental volume is ≈100 liters.

## References

1. V. Chazal et al., Astropart. Phys. **9** (1998) 163.
2. P. Di Stefano et al., Astropart. Phys. **14** (2000) 329.
3. see, e.g., R.J. Gaitskell, these proceedings, and references therein
4. P.F. Smith et al, Phys. Rep. **307** (1998) 275.
5. A. Benoit et al., Phys Lett. B **479** (2000) 8.
6. X.F. Navick et al., Nucl. Instr. and Meth. A **444** (2000) 361.
7. R. Bernabei et al., Phys. Lett. B **450** (1999) 448.
8. R. Abusaidi et al., Phys. Rev. Lett. **84** (2000) 5699.
9. L. Baudis et al., Phys. Rev. D **59** (1999) 022001.